\documentclass[fleqn]{mn2e}                                                                             

\usepackage{epsfig}
\usepackage{graphics}
\usepackage{amsmath}

\title{Carbon Monoxide Depletion in Orion B Molecular Cloud Cores}
\author[D. Savva, L.T. Little, R.R. Phillips, A.G. Gibb]
       {D. Savva$^1$\, L.T. Little$^1$, R.R. Phillips$^2$, A.G. Gibb$^3$\\
        \small $^1$Department of Electronics, University of Kent at
        Canterbury,  Canterbury, Kent, CT2 7NT, UK.\\
        \small $^2$Joint Astronomy Centre, 660 North A'ohoku Place,
        University Park, Hilo, HI 96720, USA. \\ 
        \small $^3$Department of Astronomy, University of Maryland,
        College Park, MD 20742, USA.} 

\date{Accepted 2003......
      Received 2002.....;
      in original form 2002.....}

\pagerange{\pageref{firstpage}--\pageref{lastpage}}
\pubyear{2003}

\begin{document}

\maketitle

\label{firstpage}

\begin{abstract}
  
  We have observed several cloud cores in the Orion B (L1630)
  molecular cloud in the 2--1 transitions of C$^{18}$O, C$^{17}$O and
  $^{13}$C$^{18}$O. We use these data to show that a model where the
  cores consist of very optically thick C$^{18}$O clumps cannot
  explain their relative intensities. There is strong evidence that
  the C$^{18}$O is not very optically thick. The CO emission is
  compared to previous observations of dust continuum emission to
  deduce apparent molecular abundances. The abundance values depend
  somewhat on the temperature but relative to `normal abundance'
  values, the CO appears to be depleted by about a factor of 10 at the
  core positions. CO condensation on dust grains provides a natural
  explanation for the apparent depletion both through gas-phase
  depletion of CO, and through a possible increase in dust emissivity
  in the cores. The high brightness of HCO$^{+}$ relative to CO is
  then naturally accounted for by time-dependent interstellar
  chemistry starting from `evolved' initial conditions.  Theoretical
  work has shown that condensation of H$_{2}$O, which destroys
  HCO$^{+}$, would allow the HCO$^{+}$ abundance to increase while
  that of CO is falling.

\end{abstract}


\section{Introduction}

There is considerable evidence that protostellar collapse occurs from
an interstellar envelope (scale $\sim$10$^{4}$AU) onto, most likely, a
circumstellar disc (scale $\sim$100 AU) . For a long time the collapse
is likely to be isothermal and at a low temperature (about 10 K). This
temperature is lower than the freeze-out temperatures of common
interstellar molecules (CO 15--17 K, NH$_{3}$ 50--60 K, H$_{2}$O 90 K;
see Nakagawa 1980), and, as collapse proceeds, the freeze-out
timescale becomes shorter than the collapse timescale. The expectation
is for a long isothermal phase at $\sim$10 K when the molecules
freeze-out onto grains followed by re-heating when they come off
again. These processes have been modelled by, inter alia, Nejad,
Williams \& Charnley (1990), Rawlings et al. (1992), Bergin \& Langer
(1997) and Charnley (1997).

There is observational evidence for depletion in cores (by about an
order of magnitude) in molecules such as CO and CS (e.g. Bergin et al.
2001; Tafalla et al. 2002). Most of these results have come from
observations of nearby molecular cloud cores in which low-mass stars
are forming in small groups (e.g. Jorgensen, Sch\"oier \& van Dishoeck
2002). In the more distant Orion region where our targets lie, NGC2024
(in L1630) contains several compact cores which show evidence for
depletion of CO and other molecules (Mauersberger et al. 1992).

The interpretation as depletion has been criticised, however, by
Chandler and Carlstrom (1996), who have provided evidence that some of
the molecular gas is hot.  Their criticisms include: (a) the dust
opacity law in the cores will be non-standard, (b) dust emission may
become optically thick at the short submm wavelengths -- giving an
`artificially low' flux which is interpreted as a low temperature in
an `optically thin' interpretation, and (c) the cores may contain
unresolved components which are highly optically thick in molecular
lines - but thin in submillimetre dust emission. Thus the true
molecular abundance may be much higher. Mangum et al. (1999) also
provide evidence that the gas in the cores of NGC2024 is hot ($>$40 K)
which would accommodate a normal abundance for CO.

One problem with NGC2024, however, is that it is near strong external
heat sources and HII emission and it is not clear that the line
emission from the molecular gas arises from within the cores from
which the dust emission emanates.

In this paper we report new observations of carbon monoxide isotopomers
C$^{18}$O, C$^{17}$O and $^{13}$C$^{18}$O from selected positions near and
around cold cores in the same molecular cloud, L1630. These positions have been
chosen with reference to previously published maps of HCO$^{+}$, C$^{18}$O and
dust continuum emission from the condensations LBS23 (also known as HH24--26),
LBS17, and LBS18 (e.g. Gibb et al. 1995; Gibb \& Little 1998, 2000; Phillips,
Gibb \& Little 2001. These references will be quoted as GLHL, GL98, GL00, and
PGL respectively). GL98 and GL00 deduced that there was depletion in several
cores. The aim of this work is to confirm its significance and to evaluate its
extent. We seek to determine optical depths from the CO isotopomers and use the
previously published dust emission data together with a range of possible
emission laws to examine the relative abundance.

In Sections 2 and 3 we justify our need to observe isotopomers as rare as
$^{13}$C$^{18}$O. In Section 4 we describe the observations, which we
analyse in Section 5 to deduce the depletion in and around the cores.
The chemical significance of the results is outlined in Section 6.

\section{Proving depletion?}

\begin{figure} \psfig{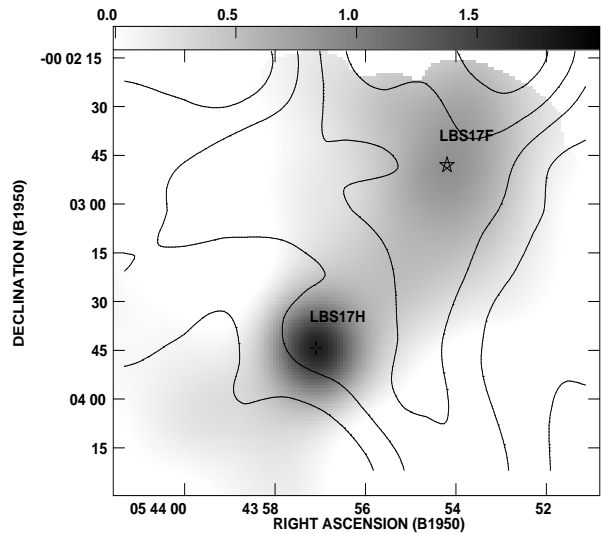}
\caption{LBS17 (22 arcsec beam). Contours are C$^{18}$O J=2$\rightarrow$1
emission integrated from LSR=7 km s$^{-1}$ to 13 km s$^{-1}$. Contours at 3, 4,
5, 6 and 7 K-km s$^{-1}$, and peak intensity of 7.60 K-km s$^{-1}$. Greyscale
is 850-$\mu$m emission with a peak flux of 1.76 Jy/beam.}  \end{figure}

\begin{figure} \psfig{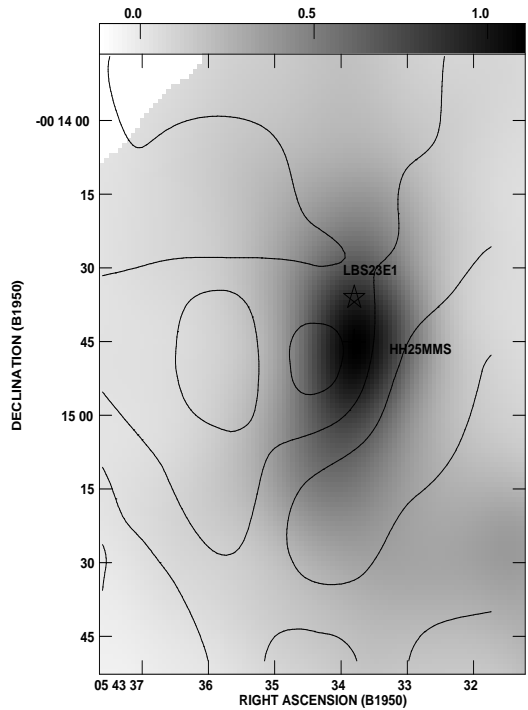}
\caption{HH25MMS (22 arcsec beam). Contours are C$^{18}$O J=2$\rightarrow$1
emission integrated from LSR=7 km s$^{-1}$ to 13 km s$^{-1}$. Contours at 1.65,
3.3 and 4.95 K-km s$^{-1}$, and peak intensity of 5.51 K-km s$^{-1}$. Greyscale
is 850-$\mu$m emission with a peak flux of 1.10 Jy/beam }  \end{figure}

\begin{figure} \psfig{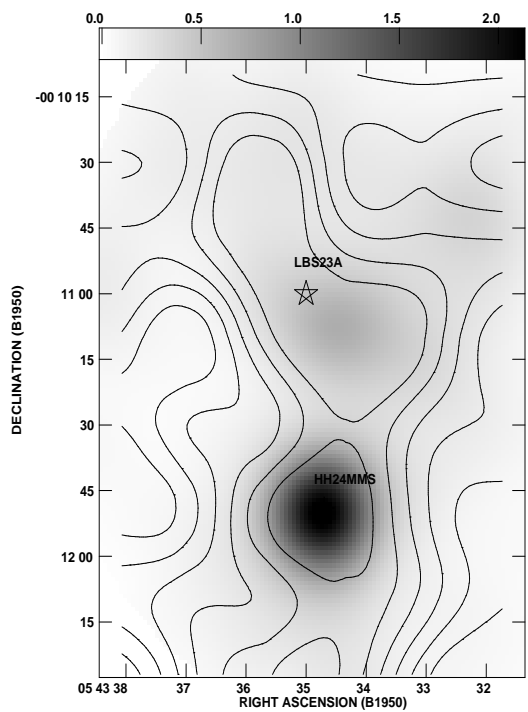}
\caption{HH24MMS (22 arcsec beam). Contours are C$^{18}$O J=2$\rightarrow$1
emission integrated from LSR=7 km s$^{-1}$ to 13 km s$^{-1}$. Contours at 4.4,
3.2, 4.0, 4.8, 5.6 and 6.4 K-km s$^{-1}$, and peak intensity of 7.66 K-km s$^{-1}$.
Greyscale is 850-$\mu$m emission with a peak flux of 2.21 Jy/beam} 
\end{figure}

Maps of two of the largest cores in the Orion B molecular cloud, LBS17
and LBS23, made in high excitation HCO$^{+}$
and C$^{18}$O (J=2$\rightarrow$1) lines , show striking differences
between the structures traced by the two molecules. Particularly
apparent is the enhanced brightness of $J$=3$\rightarrow$2 HCO$^+$
relative to $J$=2$\rightarrow$1 C$^{18}$O in the protostellar core
LBS17-H (Fig. 2 of GL00), and of $J$=4$\rightarrow$3 HCO$^+$ to
$J$=2$\rightarrow$1 C$^{18}$O in HH25MMS (Fig. 1 of GL98).
Figs. 1, 2 and 3 show the relation between the C$^{18}$O and dust
emission. C$^{18}$O emission is surprisingly weak in some of the
clumps which stand out prominently in the other species. An
interpretation assuming a simple source structure, optically thin
C$^{18}$O, and the dust emissivity law proposed by Hildebrand (1983),
implies widespread reductions of the C$^{18}$O abundance in these
clumps by factors of 13 to 57 compared to what we take to be its
canonical value of 2$\times$10$^{-7}$ (GL98 and Frerking, Langer \&
Wilson 1982).  Most of the clumps appear to be bound objects of
several solar masses (for LBS23 at least).

GL98 consider other ideas to explain the apparent reduction in
abundance for the clumps in LBS23. Abnormal dust properties do not
appear to be required because dust-emission-derived masses for the
clumps are in good agreement with their virial masses.

Another possibility is the absorption of C$^{18}$O emission from
intervening low excitation temperature gas. If this were the case then
the effect would be seen as strong self-absorption in the optically
thicker CO lines. However, there is no evidence for such
self-absorption in the CO spectra (GH93). Since the observed antenna
temperatures measure the excitation temperature where the optical
depth becomes equal to unity, a comparison of the CO and C$^{18}$O
antenna temperatures would then suggest that the cloud cores are
externally heated (see e.g. GL98).

If the C$^{18}$O emission arises in optically thick, unresolved
sub-clumps the need for depletion might be avoided as all the
C$^{18}$O might not be detected.  Accordingly we have made careful
observations of weak isotopomers, including C$^{18}$O, C$^{17}$O and
$^{13}$C$^{18}$O, to seek to eliminate a sub-clump model. To
illustrate how this can be achieved we consider a simple two-component
model in the next section.

\section{Optically thick clumping model}

\begin{table*}
\begin{center}
\caption{Line optical depth and brightness (K) predictions for the
  `sub-clump' and `depletion' models, and observed values, for
  HH25MMS.} 
\begin{tabular}{lcccccc}
\hline
                                           
        &\multicolumn{4}{c}{Sub-clump model}   &Depletion      &\\

Molecule        &Sub-clump     & $T^*_{\rm A_1}$  (K)  &$T^*_{\rm A_2}$  (K)  &Total  &model  &Observed       \\
        & optical depth        &  sub-clump    &Envelope     &  (K)
        & (K)      & (K)       \\

\hline             

C$^{18}$O       &120    &0.34   &1.1    &1.4    &1.4    &1.3    \\
C$^{17}$O       &22     &0.34   &0.2    &0.55   &0.25   &0.28   \\
$^{13}$C$^{18}$O  &2      &0.34   &0.02   &0.35   &0.025        &$<$0.08        \\

\hline
\label{clumps}
\end{tabular}
\end{center}
\end{table*}

In Section 5 of GL98 the authors describe a two-component model
(`envelope' and `sub-clump') where the envelope (component 1) is
optically thin (depth $\tau_{\rm 1}$, linewidth $\Delta v_{\rm 1}$) in the less abundant CO isotopomers, such as C$^{18}$O,
and the sub-clump material (component 2) is optically thick (depth $\tau_{\rm
2}$, linewidth $\Delta v_{\rm 2}$) in these
same lines. Extending their analysis it is also possible to define an \emph{apparent} depletion
factor which describes how the presence of the optically thick clump
mimics depletion within the beam in the simplistic interpretation as a
single component. The CO
abundance for such a source would be 
proportional to $(T^*_{\rm A1} \Delta v_{\rm 1}$ + $T^*_{\rm A2}
\Delta v_{\rm 2} \tau_{\rm 2})$.
However, in the standard optically thin analysis, not knowing $\tau_2$, we would assume the CO abundance is
proportional to $(T^*_{\rm A1} \Delta v_{\rm 1}$ + $T^*_{\rm A2}
\Delta v_{\rm 2})$.
We will thus deduce CO abundances that are too low by a ratio
\begin{equation}
A = \frac{T^*_{\rm A1} \Delta v_1 + T^*_{\rm A2} \Delta v_2}{T^*_{\rm A1} \Delta v_1 +
  T^*_{\rm A2} \Delta v_2 \tau_2} = \frac{R_{12} + 1}{R_{12} + \tau_2} 
\label{eqna}
\end{equation}
where $R_{12}$ is defined in GL98 as the ratio of the integrated
intensities of components 1 and 2.
The optical depth of the sub-clump component $\tau_2$ in the denominator causes
$A$ to be less than unity, and demonstrates how an optically thick
sub-clump can mimic depletion. If we assume that the abundance is in fact normal then we can use our
model along with estimates of $R_{\rm 12}$ and $A$ to deduce
information concerning the nature of the sub-clumps.

For example, the clump HH25MMS (GLHL, GL98 and PGL) stands out very
prominently in dust continuum and 4$\rightarrow$3 HCO$^{+}$ emission
but it is barely detectable as an independent source on the
north-south ridge of 2$\rightarrow$1 C$^{18}$O emission. The
integrated intensity on the ridge is typically 2.5 K\,km\,s$^{-1}$
while HH25MMS adds about 0.8 K\,km\,s$^{-1}$ to it. Thus $R_{12} =
2.5/0.8 = 3.1$. Also the apparent depletion factor, $A$, is about 1/30
(according to GL98).  Hence, from equation \ref{eqna}, $\tau_{\rm 2}$ = 120.

GL98 showed that the beam filling factor of the sub-clumps was given by
\begin{equation}
f_c = \frac{\tau_1 \Delta v_1}{R_{12} \Delta v_2 + \tau_1 \Delta v_1}
\end{equation}

\noindent The 4$\rightarrow$3 HCO$^{+}$ emission is closely correlated with the
continuum dust emission, rather than the C$^{18}$O, but has a
linewidth similar to the C$^{18}$O so $\Delta v_{\rm 1} = \Delta
v_{\rm 2}$. Also for HH25MMS $\tau_{\rm 1} \sim$ 0.2 so that the
sub-clump filling factor is $f_{\rm c}$ = 0.06. These would be sub-clumps
with a very high C$^{18}$O optical depth and a very small beam filling
factor in the 22 arcsec JCMT beam.

We shall assume that the JCMT beam, of radius \emph{R}, contains
\emph{n} identical sub-clumps of radius \emph{r}. Then $f_{\rm c} =
n(r/R)^{2}$. We know $n \geq 1$ since HH25MMS is resolved. It follows
that $r \leq 2.4$ arcsec or 1000 AU at the distance of L1630 (400 pc).
Since the apparent depletion is high, most of the mass within the beam
at HH25MMS is in the sub-clumps. GLHL quote a mass of 8 M$_\odot$
in the core of HH25MMS. The sub-clump mass is then $\sim
(8/n)$ M$_\odot$ and its density is greater than 2.5$\times$10$^{8}$
cm$^{-3}$. The density of the envelope material is much lower. Its
mass is 1/30th that of the clumps, and taking the mass to be within a
beamwidth gives a density 5$\times$10$^{4}$ cm$^{-3}$. In summary, if
we seek to retain normal abundances by assuming sub-clumping it would
appear that the sub-clumps must have radii less than 1000 AU and
densities greater than 2.5$\times$10$^{8}$ cm$^{-3}$.

As described by GL98, the most obvious problem with the `sub-clump'
interpretation seems to lie in the fact that the HCO$^{+}$ intensity
follows the submillimetre dust continuum emission well (rather than
the envelope observable in C$^{18}$O) yet its brightness temperature
is so high that it must have a beam filling factor near unity, which
is much greater than that deduced for the sub-clumps from C$^{18}$O.
Nonetheless it is important to test the structure within the clumps
either via interferometric observations or beam-matched observations
of weaker CO isotopomers.

We may predict C$^{17}$O line intensities, assuming $\tau_{\rm
  2}$=120. The component 1 intensity should decrease by a factor of
5.4 compared to C$^{18}$O, since it is optically thin, but that from
component 2 will be unaltered as its optical depth will still be
extremely high (=120/5.4). Referring to Table \ref{clumps}, the intensity will be
\begin{equation}
T^*_{\rm A} = T^*_{\rm A1} + T^*_{\rm A2} = 0.34 \,{\rm K} + 0.2 \,{\rm K} = 0.54
\,{\rm K} 
\end{equation}
so the ratio of the C$^{18}$O to C$^{17}$O intensities will be 2.6.
This might easily be misinterpreted as indicating that the C$^{18}$O
was just slightly optically thick. Observations of even rarer
isotopomers -- so that the optically thin `envelope' contribution
becomes negligible while the optically thick core component still
remains -- are required. Table \ref{clumps} gives predicted line intensities for
the `depletion' and `sub-clump' models. It can be seen that
observations of $^{13}$C$^{18}$O are clearly capable of
distinguishing between them.

\section{Observations}

\subsection{CO observations}

Observations of the J=2$\rightarrow$1 transitions of C$^{18}$O, C$^{17}$O and
$^{13}$C$^{18}$O were carried out over the period of 1998 February 16--28 at the
James Clerk Maxwell Telescope, using the common user SIS receiver A2 (Davies et al.
1992) and the digital autocorrelation spectrometer DAS. The original mixer in A2 had
been replaced by one from NRAO. The C$^{17}$O observations were made assuming a rest
frequency of 224.714 GHz, the C$^{18}$O observations at 219.560 GHz and the
$^{13}$C$^{18}$O observations 209.419 GHz. Key positions were selected from our
previous maps of the cores, and, in addition, a new map of C$^{18}$O in LBS18 was
made. The positions chosen are listed in Table \ref{pos}. The positions of cores were
observed, as well as a selection of references off core where there was sign of
continuum emission. 

The pointing and antenna temperature were measured with reference to
the nearby source OMC1. The pointing was consistent typically to
within about 3 arcsec. On 16th and 17th February the apparent antenna
temperatures of lines from the frequently observed calibrating source
OMC1 varied with time when observing the C$^{18}$O and
$^{13}$C$^{18}$O lines. The intensity of the C$^{18}$O line varied
systematically with time between 0.51 and 0.80 of that of the
Observatory reference spectrum for OMC1 C$^{18}$O, $T_{\rm A}^{*}$ =
7.5 K. The C$^{18}$O intensities we quote have been systematically
corrected for this variation, which brought three of the observations
into agreement with similar ones described by GL98.

There was no appropriate Observatory reference for that part of the
passband containing $^{13}$C$^{18}$O or C$^{17}$O. Although
$^{13}$C$^{18}$O was not clearly discernible for OMC1 on 16th February
the intensities of several bright lines in the passband containing it
varied systematically with time between 0.42 and 0.64 of their value
on 18th and 19th February. It has been assumed that the values for
18th and 19th February are correct, and our $^{13}$C$^{18}$O
intensities scaled on this assumption. The C$^{17}$O line observed in
OMC1 showed no significant variation between 17th, 18th and 20th
February, yielding $T_{\rm A}^{*}$ = 1.9$\pm$0.1 K. This value has
been assumed correct, and no scaling applied to our observed C$^{17}$O
intensities.

\begin{table}
\begin{center}
\caption[]{Positions studied in this investigation (B1950).}
\begin{tabular}{lccccccc}
\hline
Source  &\multicolumn{3}{c}{RA} &\multicolumn{3}{c}{DEC}        & Core? \\
        & h     & m     &s      & $^\circ$        &\arcmin & \arcsec       &       \\
\hline
HH24MMS  &05     &43     &34.7   &--00    &11     &49.0   & yes   \\
HH25MMS &05     &43     &33.8   &--00    &14     &45.0   & yes   \\
LBS17H  &05     &43     &57.1   &--00    &03     &44.3   & yes   \\
LBS18S  &05     &43     &54.2   &+00    &18     &23.0   & yes   \\
\\
LBS23E1 &05     &43     &33.8   &--00    &14     &36.0   & no    \\
LBS17F  &05     &43     &54.2   &--00    &02     &48.0   & no    \\
LBS18A  &05     &43     &55.2   &+00    &18     &57.0   & no    \\
LBS18B  &05     &43     &54.7   &+00    &18     &48.0   & no    \\
LBS23A  &05     &43     &35.0   &--00    &11     &00.0   & no    \\
\hline
\label{pos}
\end{tabular}
\end{center}
\end{table} 

An observation of CO in OMC1 on 17th February yielded a line of
brightness 0.9 of the observatory reference value. The reason for the
time variation of C$^{18}$O and $^{13}$C$^{18}$O line intensities on
16th and 17th February has not been established. Based on the authors'
experience, possibilities might include incorrect adjustment of one of
the telescope mirrors, problems in SIS receiver calibration (see, e.g.
Davies et al. 1992) or uncertainty in receiver sideband calibration,
though the latter might not be expected to vary with time. The
receiver was tuned for double sideband performance and it is expected
that the sideband ratio should be no worse than 1.2. Changing the LO
frequency to observe the same C$^{18}$O line in both sidebands on the
17th, when the observed strength was low (though not a measurement of
the sideband ratio) produced very similar line intensities, tending to
suggest that the sideband ratio was not responsible.

\begin{figure} \psfig{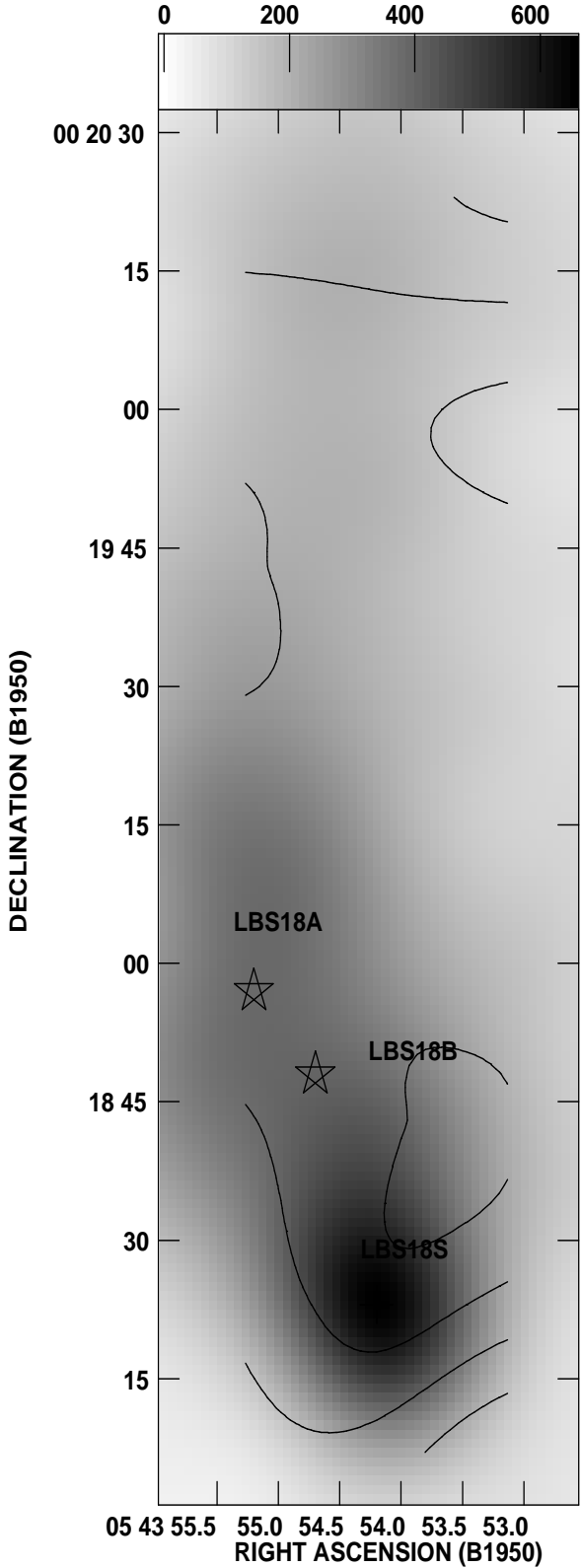}
\caption{LBS18 (22 arcsec beam). Contours are C$^{18}$O J=2$\rightarrow$1
emission integrated from LSR=7 km s$^{-1}$ to 13 km s$^{-1}$. Contours at 1.25,
1.75, 2.25, 2.75 and 3.25 K-km s$^{-1}$, and peak intensity of 3.35 K-km s$^{-1}$. Greyscale is
850-$\mu$m emission with a peak flux of 0.65 Jy/beam.} 
\label{lbs18comap}
\end{figure}

\begin{figure} \epsfig{figure=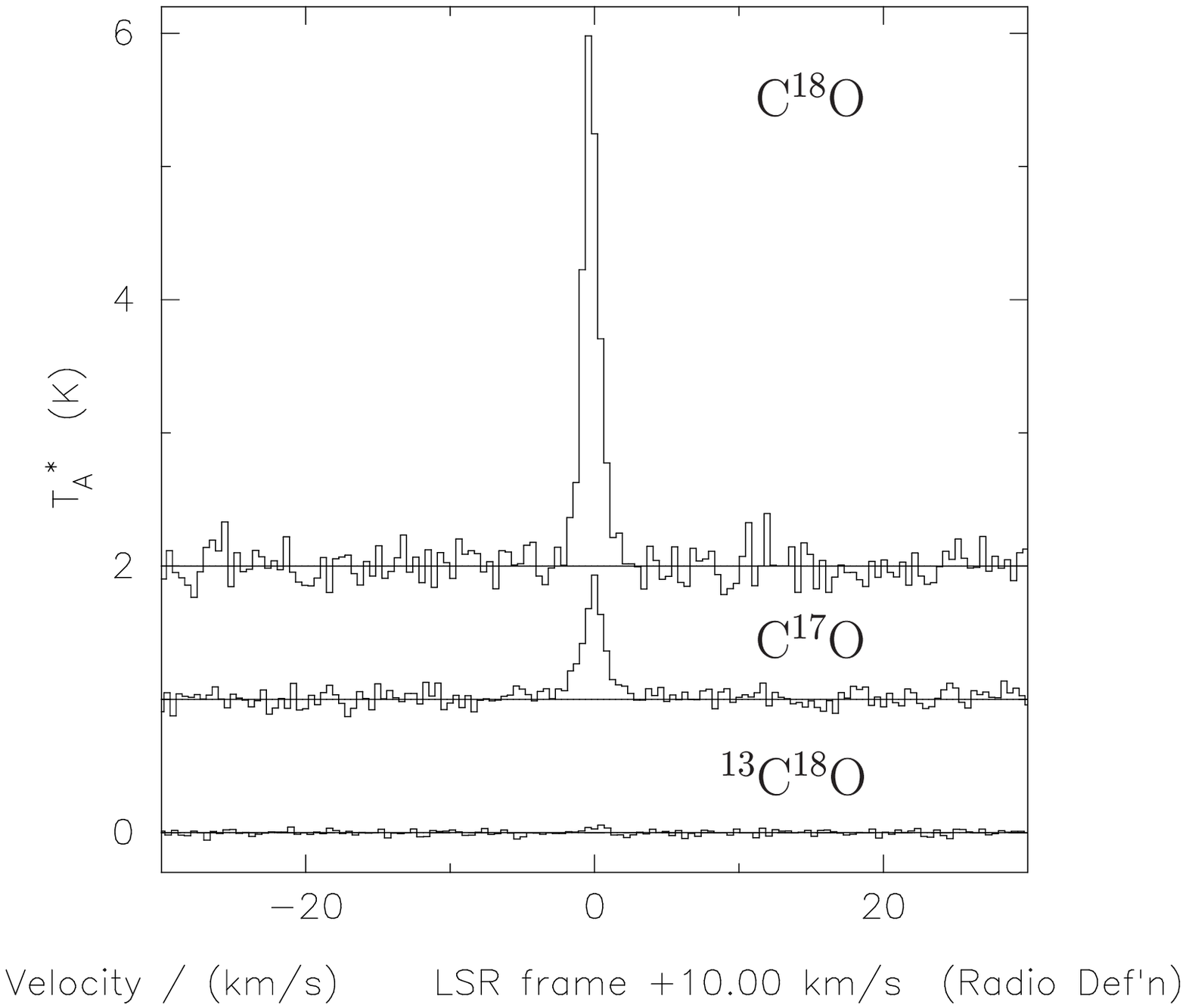, height=8cm, width=8cm}
\caption{HH24MMS: C$^{18}$O, C$^{17}$O and
$^{13}$C$^{18}$O spectra at the core position (RA=05$^h$43$^m$34.7$^s$
DEC=-00$^{\circ}$11{\arcmin}49.0{\arcsec}).}   
\label{hh24spec} 
\end{figure}

Fig. \ref{lbs18comap} shows the new map of the LBS18 region along
with the 850 $\mu$m map from PGL, and Figs. \ref{hh24spec} to \ref{lbs18spec} show
the new spectra obtained for HH24MMS, HH25MMS, LBS17H and LBS18S.

\subsection{Continuum observations}

The continuum observations used in this paper were made using the
submillimetre bolometer array receiver SCUBA and are described in PGL.
For the purposes of this paper we have smoothed their 850$\mu$m maps
to the same resolution as that of our CO data, i.e. 22 arcsec.

\begin{figure} \epsfig{figure=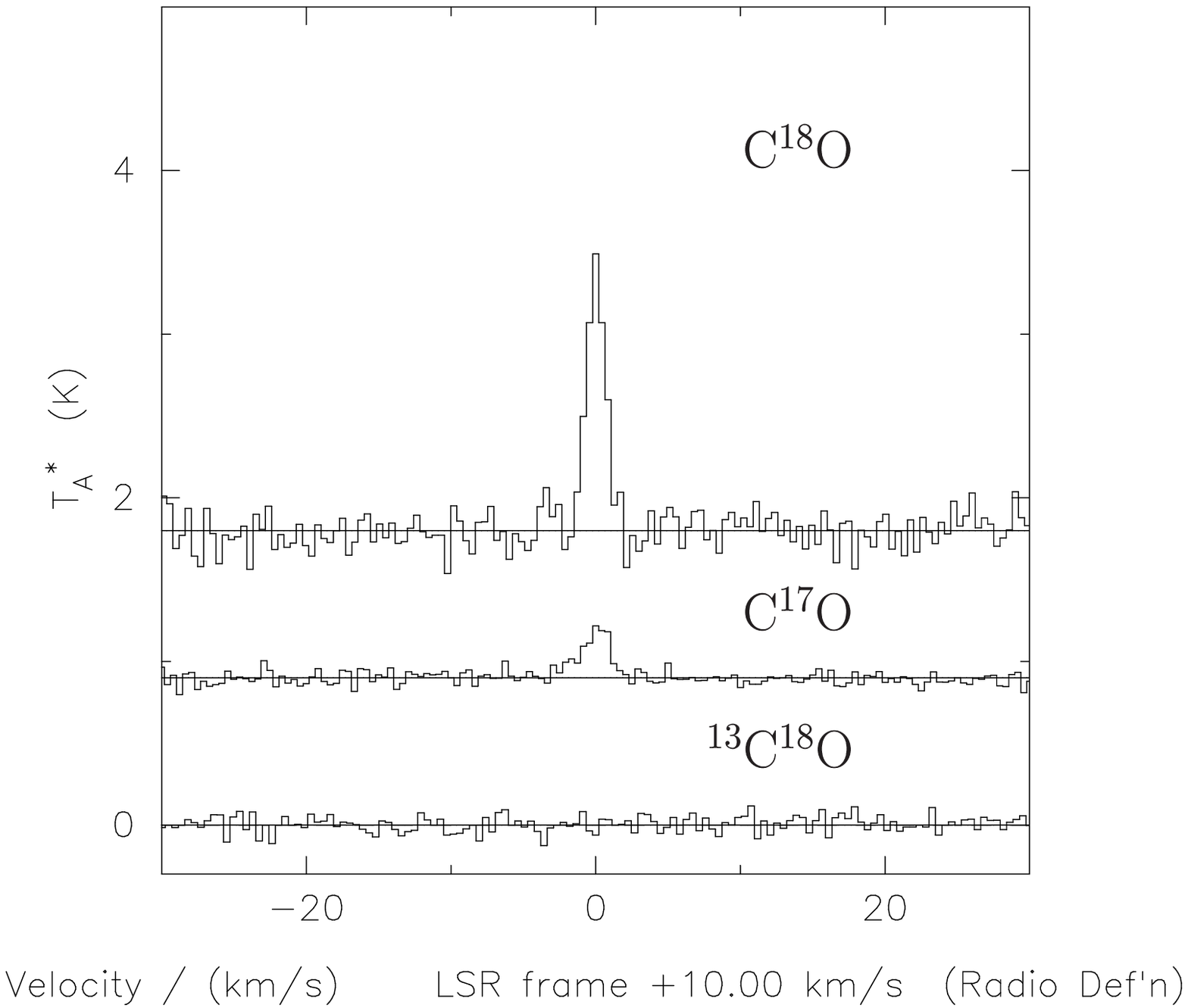, height=8cm, width=8cm}
\caption{HH25MMS: C$^{18}$O, C$^{17}$O and
$^{13}$C$^{18}$O spectra at the core position (RA=05$^h$43$^m$33.8$^s$
DEC=-00$^{\circ}$14{\arcmin}45.0{\arcsec}).} 
\end{figure}
\begin{figure} \epsfig{figure=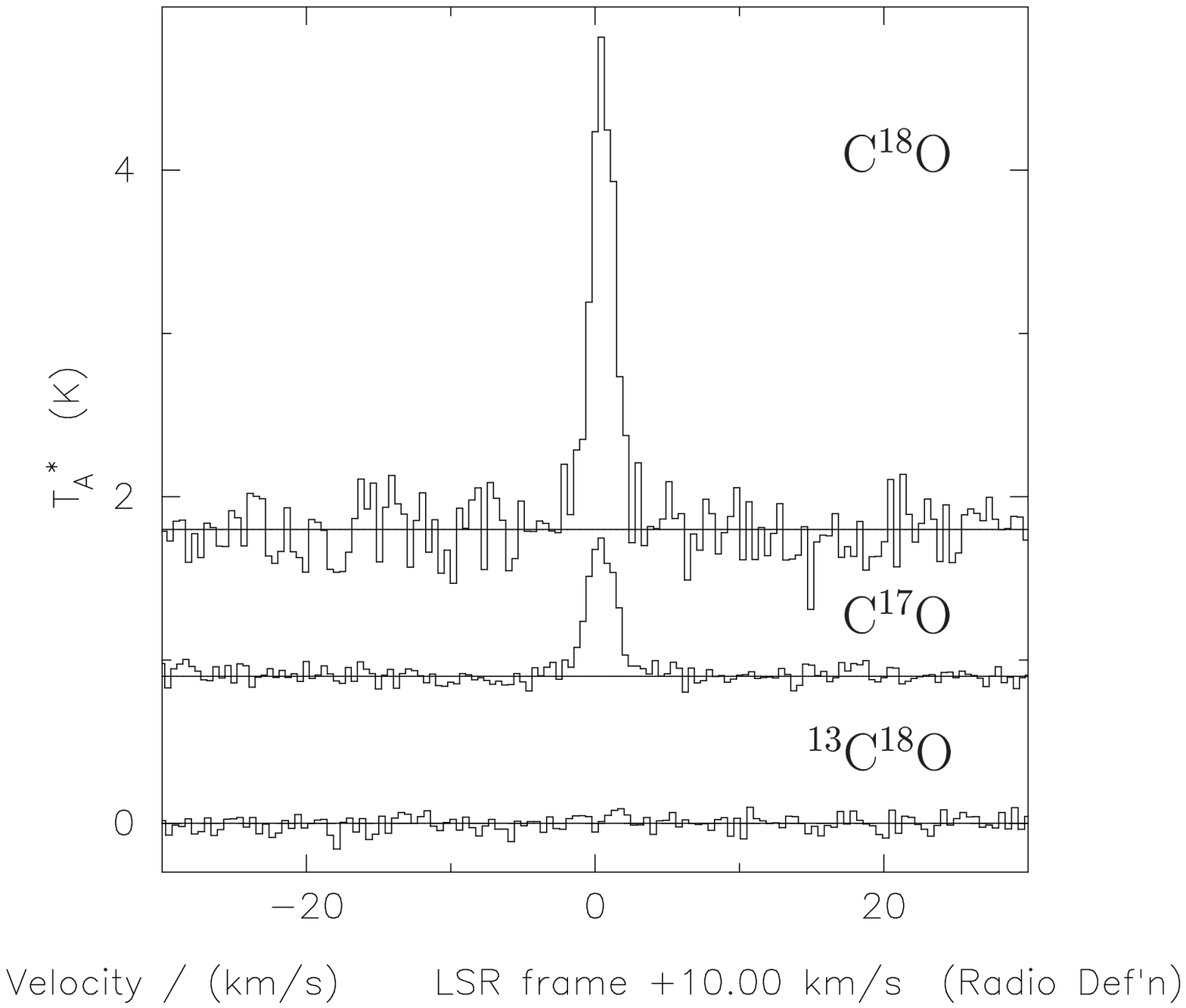, height=8cm, width=8cm}
\caption{LBS17H: C$^{18}$O, C$^{17}$O and
$^{13}$C$^{18}$O spectra at the core position (RA=05$^h$43$^m$57.1$^s$
DEC=-00$^{\circ}$03{\arcmin}44.3{\arcsec}).} 
\end{figure}
\begin{figure} \epsfig{figure=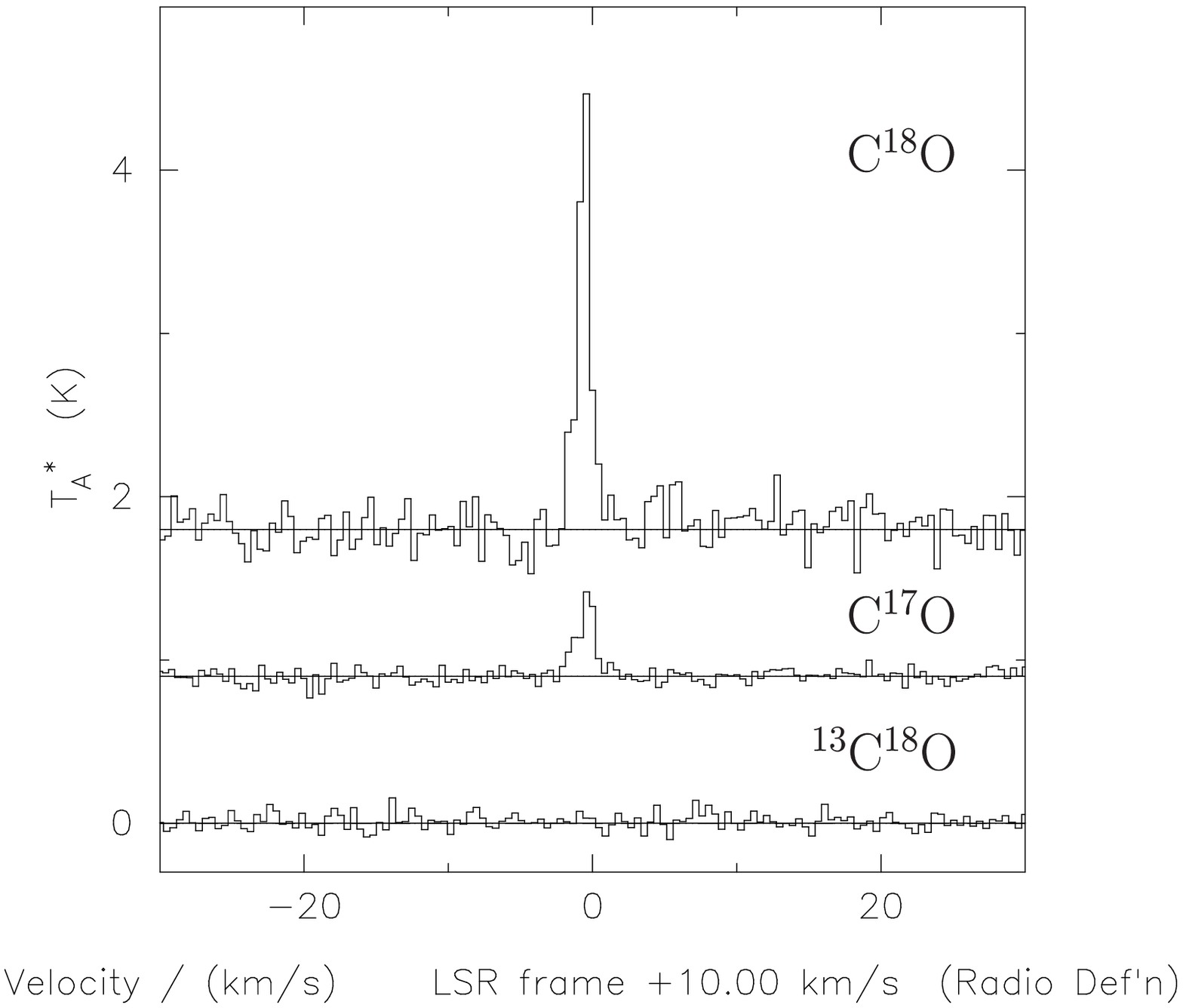, height=8cm, width=8cm}
\caption{LBS18S: C$^{18}$O, C$^{17}$O and
$^{13}$C$^{18}$O spectra at the core position (RA=05$^h$43$^m$54.2$^s$
DEC=+00$^{\circ}$18{\arcmin}23.0{\arcsec}).} 
\label{lbs18spec}
\end{figure}

\section{Results And Analysis}
Table \ref{intens} contains the data taken from the new observations.
The antenna temperatures, $T_{\rm A}^*$, derived from the CO
observations for all three species at each of the positions are shown.
The observations were made with a channel spacing and noise bandwidth
of 0.45 kms$^{-1}$. Four channels were averaged to give the
temperatures quoted in Table \ref{intens}. (Note that the spectra are
show in the original spectral resolution of 0.45 km\,s$^{-1}$.) The
integrated intensity $I_{\rm int}$(K\,km\,s$^{-1}$), which is derived
from the antenna temperature ($I_{\rm int}$=$\int T_{\rm A}^*dv$,
integrated over $\pm$1.79 kms$^{-1}$), for each species is also shown.

The Astronomical Image Processing Software ({\scshape aips\/}) package
was used to analyse the continuum data. The 850 $\mu$m and 450 $\mu$m
continuum emission maps were convolved to an effective resolution of
22 arcsec. The fluxes $F$ (Jy) for each position listed in Table
\ref{pos} were then taken from the convolved maps and are listed in
Table \ref{intens}. The uncertainty in these values is 0.02--0.03 Jy.

\begin{table*}
\begin{center}
\caption[]{C$^{18}$O, C$^{17}$O and $^{13}$C$^{18}$O line temperatures
  and integrated intensities, and 850 $\mu$m flux measurements at each
  position. Note that the noise levels quoted are derived from spectra
  binned by four channels and are thus a factor of 2 lower than might
  be expected from a direct comparison with Figs. \ref{hh24spec} to
  \ref{lbs18spec}.}
\begin{tabular}{|l|cc|cc|cc|c|}
\hline
        &\multicolumn{2}{|c|}{C$^{18}$O}                &\multicolumn{2}{|c|}{C$^{17}$O}                &\multicolumn{2}{|c|}{$^{13}$C$^{18}$O}   &\multicolumn{1}{c|} {850 $\mu m$}   \\

Source  &$T_{\rm {A}^*}$  &$I_{\rm int}$    &$T_{\rm {A}^*}$  &$I_{\rm
        int}$    &$T_{\rm {A}^*}$  &$I_{\rm int}$    &$F$        \\ 
        &(K)    &(K km s$^{-1}$)        &(K)    &(K km s$^{-1}$)
        &(K)    &(K km s$^{-1}$)        &(Jy)   \\ 
\hline
HH24MMS &2.8$\pm$0.03   &5.8$\pm$0.07  &0.67$\pm$0.01  &1.75$\pm$0.03
        &0.057$\pm$0.020 &0.09$\pm$0.03  &4.76   \\ 

HH25MMS &1.3$\pm$0.03   &2.5$\pm$0.09  &0.28$\pm$0.01  &0.80$\pm$0.02
&$<$0.080$\pm$0.045       &$<$0.02$\pm$0.05       &2.36   \\ 

LBS17H  &2.6$\pm$0.03   &6.0$\pm$0.07  &0.64$\pm$0.01  &1.80$\pm$0.03
&0.088$\pm$0.048         & 0.16$\pm$0.06  &3.79   \\ 

LBS18S  &1.5$\pm$0.03   &3.0$\pm$0.16  &0.37$\pm$0.01  &0.80$\pm$0.03
&$<$0.072$\pm$0.048       &$<$0.02$\pm$0.06       &1.34   \\ 

\\

LBS23E1 &1.8$\pm$0.06   &3.4$\pm$0.15  &0.30$\pm$0.02  &0.80$\pm$0.05
&---    &---    &1.97   \\ 

LBS17F  &3.9$\pm$0.06   &7.3$\pm$0.15  &0.90$\pm$0.02  &2.20$\pm$0.05
&---    &---    &1.85   \\ 

LBS18A  &2.6$\pm$0.06   &3.0$\pm$0.15  &0.45$\pm$0.02  &0.75$\pm$0.05
&---    &---    &0.82   \\ 

LBS18B  &3.0$\pm$0.06   &3.4$\pm$0.13  &0.42$\pm$0.02  &0.90$\pm$0.05
&---    &---    &0.85   \\ 

LBS23A  &2.6$\pm$0.10   &6.0$\pm$0.23  &---    &---
&$<$0.05$\pm$0.015       &$<$0.12$\pm$0.02       &1.08   \\ 
\hline
\label{intens}
\end{tabular}
\end{center}
\end{table*}

\subsection{Temperatures and masses}

From the fluxes at 850 $\mu$m and 450 $\mu$m we seek to derive the
temperatures and masses.  The results depend on the opacity law, which is
uncertain. Rather than seeking to justify a law we chose to investigate
the effect of varying it by comparing the different results obtained
assuming two typical but reasonably extreme examples: the laws proposed by
Hildebrand (1983) ($\beta$=2) and by Testi \& Sargent (1998)
($\beta$=1.1). These predict very similar opacities at 850 $\mu$m, but
different temperatures from 850/450 flux ratios. We accordingly derive
temperatures from both these laws, and compare them with already published
temperatures from grey body spectral fits and ammonia inversion line
observations to deduce plausible temperature limits. These are then used
with the 850 $\mu$m opacities to deduce masses.

If the absorption coefficient $\kappa(\nu)$ is written in terms of the
dust and molecular hydrogen density as
\begin{equation}
\kappa(\nu) = \kappa_{\rm d} \rho_{\rm d} = \kappa_{\rm H_{2}} n_{\rm H_{2}}
\end{equation}
the total number of hydrogen molecules $N(H_2)$ implied by
observation of a flux $F_{\rm \nu}$, from dust assumed to be optically
thin, is
\begin{equation}
N({\rm H}_2) = \frac{D^{2}c^{2}(\exp(\frac{h\nu}{kT})
  -1)}{2h\nu^{3}\kappa_{{\rm H}_{2}}} F_{\nu} 
\label{Nh2}
\end{equation}
where $D$ is the distance from Earth.

\begin{figure}
\begin{center}
\psfig{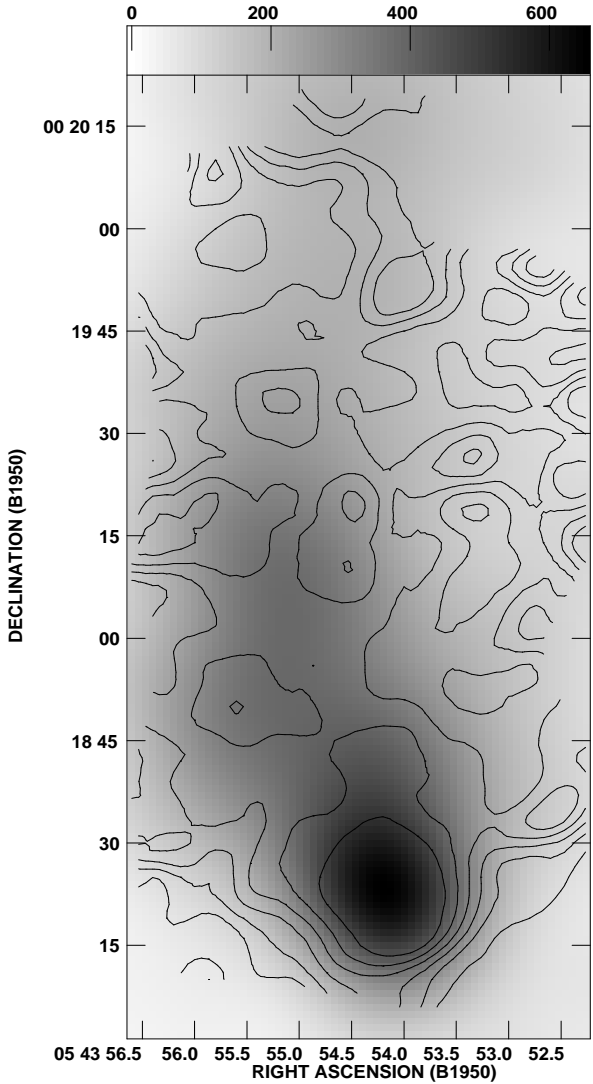}
\caption{\label{frmap} Flux-ratio map (22 arcsec) of the LBS18 region plotted over
  the 850-$\mu$m (22 arcsec) continuum map. Contours levels are at
  0.3, 0.4, 0.5, 0.6, 0.7 and 0.8 ratio, with a peak of 0.9 ratio at the dust
  peak. } 
\end{center}
\end{figure}

The temperature of the gas can be derived by forming the ratio of the
submillimetre continuum fluxes at the two wavelengths and comparing it
with that predicted for different temperatures using equation
\ref{Nh2}.  Flux-ratio maps were produced by using the {\scshape
  comb\/} task in {\scshape aips\/} to divide the 850 $\mu$m (15
arcsec resolution) maps by the 450 $\mu$m (15 arcsec resolution) maps.
The ratio, $r$(850/450), was measured at each point from these maps
and used to find the temperature. The flux-ratio map of LBS18, which
has been smoothed to a resolution of 22 arcsec, has been plotted over
the 850 $\mu$m (22 arcsec) continuum map in Fig. \ref{frmap}, where it
can be seen that the contours representing the higher values of
$r$(850/450) are coincident with the dust peaks. This may be due to
(a) a lower temperature, (b) a flatter emissivity law, or (c) the 450
$\mu$m emission being optically thick.

The resulting value of $T_{\rm d}$ for each position can be found in Table
\ref{temps} for both values of $\beta$. As expected, the temperatures are higher for
the lower value of $\beta$. The errors are derived using the calibration
uncertainties and random noise errors quoted in PGL.

Assuming the dust is optically thin can produce misleadingly low
temperatures and high derived masses if the dust is in fact optically
thick at 450 $\mu$m. However PGL used a radiative transfer program to
model the cores that was not limited by this assumption, but derived very
similar temperatures to those found here. In reality the results we find
represent an average over the beam. It is also quite likely that there are
temperature gradients within the beam, either decreasing towards the centre of
the core
(e.g., an externally heated protostellar cloud) or increasing (e.g., an
interstellar cloud heated by a central young star). If there are, then the
derived beam--averaged temperature will be weighted towards the hotter dust
which radiates more strongly. This means that the derived dust masses will
be too low (colder dust being underweighted). CO column densities are
less sensitive to temperature, so the overall effect is to derive CO
abundances which are higher than the true ones -- i.e. to make the
depletion appear less marked than it really is.

\begin{table}
\begin{center}
\caption{Temperatures (K) derived from published grey body fits, our
observations assuming $\beta$ = 1.1 and 2, and NH$_3$ observations of Harju et al.}
\begin{tabular}{|l|c|c|c|c|}

\hline
Source  &$T_{\rm gb}$   &$T_{\rm 1.1}$  &$T_{\rm 2}$    &$T_{\rm NH_{3}}$       \\
\hline
HH24MMS &20$\pm$5       &10$_{-3}^{+4}$         &7$\pm$1        &15$\pm$3       \\              
HH25MMS &17--35         &38$_{-20}^{+?}$        &13$_{-3}^{+4}$ &15$\pm$2       \\
LBS17H  &$<$20          &14$_{-5}^{+9}$         &9$_{-2}^{+3}$  &---    \\
LBS18S  &---            &14$_{-5}^{+12}$        &9$_{-2}^{+3}$  &---    \\           
\\
LBS23E1 &---    &37$_{-22}^{+?}$        &13$_{-4}^{+5}$         &---    \\
LBS17F  &---    &16$_{-8}^{+?}$         & 9$_{-4}^{+5}$         &---    \\
LBS18A  &---    &23$_{-14}^{+?}$        &11$_{-4}^{+7}$         &---    \\     
LBS18B  &---    &24$_{-14}^{+?}$        &11$_{-4}^{+6}$         &---    \\
LBS23A  &---    &37$_{-23}^{+?}$        &13$_{-5}^{+8}$         &---    \\
\hline
\label{temps}
\end{tabular}
\end{center}
\end{table}

Further information is available which can be used to constrain the temperatures.
Grey body fits to the spectra of dust continuum emission give $T_{\rm d}$=17--35 K for
HH25MMS (Gibb and Davis, 1998), $T_{\rm d}$=20$\pm$5 K for HH24MMS (Ward-Thompson et al.,
1995) and $T_{\rm d}<$20 K for LBS17H (GL00). It is also of interest to compare the
derived temperatures with those obtained from NH$_3$ inversion lines. Harju et al.
(1993) used NH$_3$(1,1) and (2,2) observations of 40 arcsec resolution to derive
the kinetic temperature of 43 star forming cores, 16 of which are associated with
the Orion L1630 and L1641 clouds. They mapped LBS23, finding temperatures of about
15K at the positions of HH24MMS and HH25MMS. These results and those from the grey
body fits are included in Table \ref{temps}. In general the kinetic temperature
shows little variation over their maps. Their results show that the average kinetic
temperature is $\sim$15.7 K for the Orion cores. Only two of their cores showed
temperatures greater than 20 K. Matthews and Little (1983) also mapped LBS23 with
130 arcsec resolution; at HH25 they found a temperature of 14 K which is similar to
that of Harju et al. As the cloud is extended and their beam contained a
substantial region surrounding the core it is likely that a temperature of 14 K is
representative of the envelope surrounding the core.

On the basis of Table \ref{temps} and the general considerations above we regard
conservative limits for the dust temperatures to be 20$\pm$5 K for HH24MMS
and HH25MMS and 15$\pm$5 K for LBS17H and LBS18S, while for the off--core
envelope positions we take $T$=15$\pm$5 K. We then use these temperatures and
the 850 $\mu$m fluxes together with the 850 $\mu$m opacity of Testi and
Sargent to deduce gas masses via equation 5. These are used in Section 5.3 to
derive relative CO abundances.

\subsection{Carbon monoxide optical depths}

We can see immediately that the observed isotope line ratios listed in
Table \ref{clumps} for HH25MMS are in much better agreement with the
predictions of the depletion model than with those for optically thick
sub-clumps. It should be noted that although optically thick
undepleted sub-clumps are unlikely to explain the apparent depletion
the existence of depleted clumps is not ruled out.

Table \ref{tau} shows the line temperatures and their ratios ($X$(18/17),
denoting the ratio of $T^*_{\rm A_{18}}$ and $T^*_{\rm A_{17}}$, the line
temperatures of C$^{18}$O and C$^{17}$O respectively). The mean value
of $X$(18/17) in the cores is 4.28; towards the other positions 6.1. The
latter value is in good agreement with the solar expectation of 5.4,
suggesting that these positions are optically thin, and that the
[C$^{17}$O]/[C$^{18}$O] ratio is normal. The former suggests that
C$^{18}$O may be slightly optically thick in the cores, but certainly
not sufficiently to discredit a former conclusion that CO is depleted
in several of the cores of LBS23 (GL98). The [C$^{17}$O]/[$^{13}$C$^{18}$O]
ratios are consistent with the solar value.
 
The above constitutes good evidence that C$^{17}$O and
$^{13}$C$^{18}$O are optically thin.

The optical depths of the three species in the individual positions
have been calculated. Firstly, the optical depth of C$^{17}$O can be
calculated using equation \ref{X12}, where $T^*_{\rm A_{17}}$ is the line
temperature of C$^{17}$O and $T^*_{\rm A_{18}}$ is that of C$^{18}$O and
$\tau_{\rm 17}$ and $\tau_{\rm 18}$ are the optical depths of
C$^{17}$O and C$^{18}$O respectively.

\begin{equation}
\frac{T^{*}_{{\rm A}_{18}}}{T^{*}_{{\rm
      A}_{17}}} = \frac{1-e^{-\tau_{18}}}{1-e^{-\tau_{17}}} =
      \frac{1-e^{-5.4\tau_{17}}}{1-e^{-\tau_{17}}} 
\label{X12}
\end{equation}

By assuming that solar values hold, $\tau_{\rm 18}$ = 5.4$\tau_{\rm
  17}$, equation \ref{X12} was solved to obtain a value for $\tau_{\rm
  17}$ at each position. The values of $\tau_{\rm 17}$ are displayed
in Table \ref{tau} with the errors calculated for each value.
Multiplying $\tau_{\rm 17}$ by 5.4 gives $\tau_{\rm 18}$ at each
position. One can also calculate \(\tau_{13,18}\), the optical depth
of $^{13}$C$^{18}$O by dividing $\tau_{\rm 17}$ by 16.6. It can be
seen from these results that all three species of CO are optically
thin in the regions where the abundances have been calculated and can
therefore be used effectively in the calculations. The negative
optical depth values in Table \ref{tau} are likely to be due to noise
errors, rather than weak maser emission.
\begin{table}
\begin{center}
\caption[]{Optical depth of C$^{17}$O at each position.}
\begin{tabular}{lccccc}
\hline
Source  & $T_{\rm A_{17}}^{*}$  & $T_{\rm A_{18}}^{*}$  & $X$(18/17)      &
$\tau_{\rm 17}$ &$\delta\tau_{\rm 17}$\\
        & (K)   & (K)                           \\
\hline
HH24MMS  & 0.67  & 2.8   & 4.2   & 0.12  & 0.02  \\
HH25MMS & 0.28  & 1.3   & 4.6   & 0.08  & 0.04  \\
LBS17H  & 0.64  & 2.6   & 4.1   & 0.13  & 0.02  \\
LBS18S  & 0.37  & 1.5   & 4.1   & 0.13  & 0.04  \\

\\
LBS23E1 & 0.30  & 1.8   & 6.0   & -0.05 & 0.02  \\
LBS17F  & 0.90  & 3.9   & 4.3   & 0.11  & 0.03  \\
LBS18A  & 0.45  & 2.6   & 5.8   & -0.03 & 0.05  \\
LBS18B  & 0.42  & 3.0   & 7.1   & -0.12 & 0.05  \\
LBS23A  & ---   & 2.6   & 7.3   & -0.13 & ---   \\
\hline
\label{tau}
\end{tabular}
\end{center}
\end{table}

\subsection{Relative abundance of CO }

For an optically thin transition the total number of molecules within
a beam $\theta$ (FWHM arcsec) from a source at D (kpc) is given by
\begin{multline}
N({\rm MOL}) = 2.5\times10^{8} \left(\frac{D}{0.4}\right)^{2}
 \left(\frac{\theta}{22}\right)^{2} \frac{(2J+1)}{J^{2}F_{J}(1-R)}
 \left(\frac{0.1}{\mu}\right)^{2} \\ 
 \times\left(\frac{55.5}{B}\right) \int T_{R}^{*}  dv \\
\label{NMOL}
\end{multline}
where $B$ is the rotational constant (GHz), $\mu$ the dipole moment
(debyes), $J$ the upper level of the transition, $F_{\rm J}$ is the
fractional population of level $J$, $\int T_{\rm R}^{*} dv$ is the
integrated line 
intensity (K\,km\,s$^{-1}$), and $R$ is given by
\begin{equation}
R = \frac{\exp(\frac{2hBJ}{kT}) - 1}{\exp(\frac{2hBJ}{kT_{\rm B}}) - 1}
\end{equation}
where $T_{\rm B}$ is the cosmic microwave background temperature (2.7
K).

The relative abundance of the CO isotopomer can be estimated by computing
$N(MOL)$ from equation \ref{NMOL} and dividing by $N({\rm H}_2)$ from equation
\ref{Nh2}. (To calculate the relative abundance, the integrated intensities in
Table \ref{intens} were divided by $\eta_{\rm B}(=0.8)$ to correct for
beam efficiency.)

The C$^{18}$O abundance, $\chi_{\rm C^{18}O}$, was calculated in the
same way as for C$^{17}$O and $^{13}$C$^{18}$O but the answers
obtained for HH24MMS, HH25MMS, LBS17H, LBS18S, and LBS17F, were
multiplied by (5.4/$X$), where $X$=$X$(18/17) (see Table \ref{tau}), to allow for
the finite optical depth of C$^{18}$O. The resulting relative abundances of C$^{17}$O,
C$^{18}$O and $^{13}$C$^{18}$O at each position can be found in Table \ref{abund}.

\begin{table}
\begin{center}
\caption{Temperatures and CO abundances}
\begin{tabular}{|l|c|c|c|c|}

\hline
Source  &$T$    &$\chi_{\rm C^{18}O}$   &$\chi_{\rm C^{17}O}$   &$\chi_{\rm ^{13}C^{18}O}$  \\
        &(K)    &$(10^{-8})$            &$(10^{-8})$            &$(10^{-10})$   \\      
\hline
HH24MMS &20$\pm$5       &2.4$_{-0.8}^{+1.1}$    &0.6$_{-0.2}^{+0.3}$  &2.8$_{-1.4}^{+1.6}$      \\                   
HH25MMS &20$\pm$5       &1.9$_{-0.7}^{+0.9}$    &0.5$_{-0.2}^{+0.2}$  &$<$1.3$_{-2.0}^{+2.0}$   \\   
LBS17H  &15$\pm$5       &2.0$_{-0.7}^{+0.8}$    &0.5$_{-0.2}^{+0.3}$  &4.1$_{-2.0}^{+2.3}$      \\          
LBS18S  &15$\pm$5       &2.9$_{-0.9}^{+1.1}$    &0.6$_{-0.3}^{+0.3}$  &$<$1.5$_{-3.0}^{+3.0}$   \\                                                     
\\
LBS23E1 &15$\pm$5       &1.7$_{-0.8}^{+0.9}$    &0.4$_{-0.2}^{+0.2}$    &---    \\           
LBS17F  &15$\pm$5       &4.8$_{-2.2}^{+2.6}$    &1.2$_{-0.5}^{+0.6}$    &---    \\         
LBS18A  &15$\pm$5       &3.6$_{-1.6}^{+1.9}$    &0.9$_{-0.4}^{+0.5}$    &---    \\                                         
LBS18B  &15$\pm$5       &3.9$_{-1.8}^{+2.1}$    &1.0$_{-0.5}^{+0.6}$    &---    \\       
LBS23A  &15$\pm$5       &5.4$_{-2.4}^{+2.9}$    &---            &$<$10.8$_{-6.1}^{+6.9}$        \\       
\hline
\label{abund}
\end{tabular}
\end{center}
\end{table}

\section{Discussion}

\subsection{Morphology}

It appears, from comparing SCUBA and line results, that the cores tend
to be found on the edges or ends of filaments. There are two kinds of
core structure: ones where the HCO$^{+}$ and dust are coincident but
C$^{18}$O is very faint or absent (LBS17H, HH24MMS), others where
HCO$^{+}$, C$^{18}$O and dust peak successively towards the cloud edge
(LBS18S and the Launhardt et al. (1996) core in LBS17 (see GL00)). The
problem is to distinguish the effects of chemistry and excitation.

\subsection{Depletion}

The canonical values for the abundance of C$^{18}$O, C$^{17}$O and
$^{13}$C$^{18}$O may be taken to be $2\times 10^{-7}$, $4.7\times
10^{-8}$ and, assuming simple proportion, $2.7\times 10^{-9}$
respectively, from Frerking et al. (1982). Dividing these values by
the corresponding abundances in each position, one can estimate the
depletion factor, $d$, for each isotopomer. Table \ref{depl} displays
these depletion factors and one can see that in all cases the CO is
depleted significantly.

Relative to the canonical values, the isotopic species appear to be
depleted by factors of about 10 on the cores, and approximately half
that in the envelopes. The values are typically less than those quoted
by GL98, the differences being due to the assumption of slightly
higher temperatures, careful matching of effective beam sizes and
correction for the optical depths of C$^{18}$O.

A natural explanation for the depletion observed in these cores is the
condensation of CO molecules on dust grains since the density
($\sim$10$^6$\,cm$^{-3}$: GLHL, GL98) is high enough to reduce the
freeze-out timescale to less than the dynamical timescale. In
addition, grain growth by coagulation is expected in such regions
which gives rise to increased dust emissivity (Ossenkopf 1993). The
consequence of these two processes is that dense collapsing cores
become very well defined in submillimetre continuum emission while
simultaneously becoming very poorly defined in C$^{18}$O emission,
such as we observe here.

A number of studies of other star-forming regions have also arrived at
the conclusion that the CO abundance is lower than the canonical
value. Studies by Kramer et al. (1999), Willacy, Langer \& Velusamy
(1998) and Caselli et al.  (1999) have found that the C$^{18}$O
abundance is reduced typically by factors ranging from a few to an
order of magnitude towards the lines of sight with highest H$_2$
column density. The study of Kramer et al.  (1999) shows that the
highest reduction in the C$^{18}$O abundance occurs in the coldest
regions. In all cases, these authors also conclude that freeze out of
CO onto cold dust grains is responsible for the observed levels of
depletion.
 
 \begin{table}
\begin{center}
\caption{Depletion factors}
\begin{tabular}{lcccc}

\hline
Source  &$T$ (K)        &$d_{\rm C^{18}O}$  &$d_{\rm C^{17}O}$  &$d_{\rm ^{13}C^{18}O}$ \\
\hline
HH24MMS &20$\pm$5       &8.5$_{-2.7}^{+4.6}$    &8.5$_{-2.6}^{+4.6}$    &9.5$_{-3.4}^{+8.9}$    \\                   
HH25MMS &20$\pm$5       &10.7$_{-3.3}^{+5.7}$   &9.2$_{-2.8}^{+5.0}$    &21.3$_{-?}^{+8.3}$     \\
LBS17H  &15$\pm$5       &9.8$_{-3.4}^{+8.1}$    &10.2$_{-3.6}^{+8.6}$   &6.6$_{-2.4}^{+6.1}$    \\ 
LBS18S  &15$\pm$5       &7.0$_{-2.3}^{+5.7}$    &8.1$_{-2.8}^{+6.6}$  &18.6$_{-?}^{+6.0}$       \\                                             
\\
LBS23E1 &15$\pm$5       &11.9$_{-4.2}^{+9.8}$   &11.8$_{-4.2}^{+9.5}$   &---            \\           
LBS17F  &15$\pm$5       &4.1$_{-1.4}^{+3.4}$    &4.0$_{-1.4}^{+3.5}$    &---            \\     
LBS18A  &15$\pm$5       &5.6$_{-2.0}^{+4.6}$    &5.3$_{-1.9}^{+4.5}$    &---            \\                                         
LBS18B  &15$\pm$5       &5.1$_{-1.8}^{+4.2}$    &4.6$_{-1.6}^{+3.8}$    &---            \\       
LBS23A  &15$\pm$5       &3.7$_{-1.3}^{+3.1}$    &---    &2.5$_{-1.0}^{+3.2}$    \\ 
\hline      
\label{depl}
\end{tabular}
\end{center}
\end{table}

Although this investigation has shown that CO appears to be depleted
in the cores, one cannot be sure that the factors are correct, as we
are calculating the abundance using averages along the line of sight.
If, for example, the cores are much colder than the envelopes, there
will be a higher depletion factor in the cores and a lower one in the
envelopes.  This is one area which needs to be studied further in
order to find a way of deriving the emission produced at the core
positions rather than averaged along the line of sight. One must also
keep in mind that the canonical values of the abundance have been
calculated using line of sight observations which may contain dust and
gas that is a lot less dense than that found in protostellar cores so
comparing the core abundances with these values may not give a true
depletion factor.

\subsection{Implications for chemical models}

Rawlings et al. (1992) show that as a protostar collapses it is
plausible that the chemistry occurring while molecules deplete out
onto dust grains leads to an increasing HCO$^{+}$/CO ratio (by up to
50), although their absolute abundances finally fall. A requirement
for this to happen is a high initial H$_{2}$O abundance such as might
result from initial conditions following a shock through already
processed material, but not from initial conditions which are of
simple atomic form. 

Observations of water have been made towards the regions encompassing
HH24MMS and HH25MMS with ISO, but these only trace a hot dense
component in shocked gas (Benedettini et al. 2000). Perhaps more
reliable but still less than ideal are the observations of NGC2024
with the Submillimeter-Wave Astronomy Satellite (SWAS) by Snell et al.
(2000).  Snell et al. estimate that the abundance of water (strictly
only the ortho species) to be $6\times 10^{-10}$, considerably lower
than expected but in agreement with the notion of ongoing and
widespread depletion by freeze out onto dust grains.

Other models have been developed which have different assumptions.
Bergin \& Langer (1997) followed chemistry during the collapse using a
gas-grain chemical code, including desorption and depletion, but
starting from an initial pristine state with elements only. The
increase in HCO$^{+}$/CO suggested by observations was not produced by
the model.  Neither was it produced by the different type of model of
Charnley (1997). 

Bergin et al. (2000) have proposed an alternative method of reducing
the CO abundance, prompted by results from SWAS. In their model CO
abundance can drop (even in warm clouds, say, 30 K) as a result of
destruction by He$^+$ with the oxygen atoms freezing out and becoming
locked in water ice. Unfortunately their discussion excludes HCO$^+$,
and it should also be noted that the CO destruction occurs on
extremely long timescales (greater than a million years), much longer
than the collapse timescales for these cloud cores.

The model of Rawlings et al. (1992) predicts an abundance for
HCO$^{+}$ in good agreement with a value we have derived for HH25MMS
($5\times10^{-10}$, GLHL). In addition, the modelling of Caselli et
al. (2002) shows that the HCO$^+$ abundance may be used to estimate
the maximum degree of depletion.  Their Model 3 (shown in their Fig.
8) also predicts a value for the HCO$^+$ abundance close to our LVG
estimate for HH25MMS. At this value (which is also at a radius close
to the beam radius for our JCMT observations) the depletion factor is
$\sim$20, in good agreement with the levels of CO depletion we derived
above.

However,it should also be noted that from our LVG modelling (GLHL) we
find that the HCO$^+$ is optically thick, and thus any decrease in the
HCO$^+$ abundance in the centre of these cores (at radii less than
$\sim$3000--4000 AU) is masked. Observations and modelling of rarer
species will help to answer this question. We are currently analysing
observations of HCO$^{+}$ and DCO$^{+}$ isotopomers on the cores to
study this phenomenon more carefully.

Since star-formation has already taken place in L1630 (e.g. Lada et
al. 1991) the initial conditions for the chemistry will not be
pristine. This supports our consideration of the model of Rawlings et
al. (1992), which starts from `evolved' conditions (including a high
abundance of water). Only this model predicts an actual increase of
HCO$^{+}$ while CO is declining. It thus provides a natural
explanation for the brightness of high excitation HCO$^{+}$ emission
relative to CO emission on cores. It thus appears that the role of
the initial conditions is very important in determining the later
behaviour of the chemistry.

\section{Conclusions}

We have used new observations of C$^{17}$O, C$^{18}$O and
$^{13}$C$^{18}$O along with existing 450 $\mu$m and 850 $\mu$m
continuum observations to study four regions in the Orion B molecular
cloud. This has allowed the determination of optical depths for the
isotopic species and the elimination of a model by which the cores
contain many optically thick sub-clumps to explain the low isotopic CO
emission.

If the dust emissivity falls within what are presently considered to be
reasonable limits, CO is depleted in both the cores and the surrounding
regions, but by a higher factor in the cores. Relative to the canonical
abundances of Frerking et al. (1982), the CO appears to be depleted by
about a factor of 10 at the core positions and a factor of 5 in the
envelopes. Although the dust observations do not allow the temperatures
to be constrained for $\beta$=1.1, we believe it unlikely, from the
NH$_{3}$ observations of Harju et al. (1993) and Matthews and Little
(1983), that temperatures will be greater than 15--20 K. It seems
unlikely that the temperatures exceed the dust derived values in the
sources.

The low abundance of CO and the brightness of HCO$^+$ emission in the
cores tends to support the model of Rawlings et al. (1992) which
suggests that the amount of HCO$^+$ in the core increases as that of
CO decreases. The HCO$^+$ abundance and observed levels of CO
depletion are also in good agreement with a recent model of molecular
ion chemistry in collapsing cores by Caselli et al. (2002).

\section*{Acknowledgments}

The authors would like to thank the JCMT staff for their efforts in
obtaining the observational data. The JCMT is operated by the Joint
Astronomy Centre on behalf of the Particle Physics and Astronomy Research
Council of the United Kingdom, The Netherlands Organization for
Scientific Research and the National Research Council of Canada. We would
also like to thank the referee for all the useful comments and
suggestions. D Savva would like to thank PPARC for the award of a
studentship.

\label{lastpage}

\end{document}